\documentclass[conference]{IEEEtran}
\IEEEoverridecommandlockouts
\usepackage{amsmath,amssymb,amsfonts}
\usepackage{placeins}
\usepackage[utf8]{inputenc}
\usepackage{graphicx}
\usepackage{epstopdf}
\usepackage{booktabs}
\usepackage{textcomp}
\usepackage{bm}
\usepackage{tcolorbox}
\usepackage{braket}
\usepackage{amsthm}
\usepackage[textsize=tiny]{todonotes}
\usepackage{yquant}
\usepackage{hyperref}
\usepackage{mathtools}
\usepackage{subcaption}
\usepackage{circuitikz}
\usepackage{tikz}
\usetikzlibrary{angles}
\usepackage{tkz-euclide}
\usepackage{float}
\usepackage{caption}
\usepackage{subcaption}
\usepackage{tabularx}
\usepackage{csvsimple}
\usepackage{xcolor}
\usepackage{comment}
\usepackage{flushend}
\usepackage{gensymb}
\usepackage{multirow}
\usepackage{svg}
\usepackage{algorithm} 
\usepackage{algpseudocode} 
\usepackage{csquotes}

\usepackage{circuitikz}
\usetikzlibrary{fit}

\definecolor{MyYellow}{HTML}{FFB000}
\definecolor{MyPink}{HTML}{DC267F}
\definecolor{MyPurple}{HTML}{785EF0}
\definecolor{MyGreen}{HTML}{029600}

\usepackage[style=ieee, maxnames=4, minnames=1, date=year, doi=false,isbn=false,backend=biber]{biblatex}

\usetikzlibrary{calc}

\def\BibTeX{{\rm B\kern-.05em{\sc i\kern-.025em b}\kern-.08em
    T\kern-.1667em\lower.7ex\hbox{E}\kern-.125emX}}

\DeclareFieldFormat
  [misc, book]
  {title}{\mkbibquote{#1}}

\AtEveryBibitem{
  \clearname{editor}%
  \clearfield{series}%
  \clearfield{isbn}%
  \clearfield{issn}%
  \clearfield{volume}%
  \clearfield{number}%
  \clearfield{pages}%
  \clearfield{url}%
  \clearfield{doi}%
}

\def\authorwidth{0.7cm}
\begin{document}
\newtheorem{exmp}{Example}

\title{Towards Equivalence Checking of\\Classical Circuits Using Quantum Computing}

\author{
	\IEEEauthorblockN{Nils Quetschlich\IEEEauthorrefmark{1}\hspace*{\authorwidth} 
 Tobias V. Forster\IEEEauthorrefmark{1}\hspace*{\authorwidth}
 Adrian Osterwind\IEEEauthorrefmark{2}\hspace*{\authorwidth}
 Domenik Helms\IEEEauthorrefmark{2}\hspace*{\authorwidth}
 Robert Wille\IEEEauthorrefmark{1}\IEEEauthorrefmark{3}}
 \IEEEauthorblockA{\IEEEauthorrefmark{1}Chair for Design Automation, Technical University of Munich, Germany}
 \IEEEauthorblockA{\IEEEauthorrefmark{2}Institute of Systems Engineering for Future Mobility, German Aerospace Center (DLR), Germany}
  \IEEEauthorblockA{\IEEEauthorrefmark{3}Software Competence Center Hagenberg GmbH (SCCH), Austria}
 \IEEEauthorblockA{\{nils.quetschlich, t.forster, robert.wille\}@tum.de 
 \hspace*{\authorwidth}
         \{adrian.osterwind, domenik.helms\}@dlr.de\\
\url{https://www.cda.cit.tum.de/research/quantum}}
}
\maketitle

\begin{abstract}
Quantum computers and quantum algorithms have made great strides in the last few years and promise improvements over classical computing for specific tasks.
Although the current hardware is not yet ready to make real impacts at the time of writing, this will change over the coming years.
To be ready for this, it is important to share knowledge of quantum computing in application domains where it is not yet represented.
One such application is the verification of classical circuits, specifically, equivalence checking. Although this problem has been investigated over decades in an effort to overcome the verification gap, how it can potentially be solved using quantum computing has hardly been investigated yet. 
In this work, we address this question by considering a presumably \mbox{straightforward} approach: Using Grover's algorithm. 
However, we also show that, although this might be an obvious choice, there are several pitfalls to avoid in order to get meaningful results. 
This leads to the proposal of a working concept of a quantum computing methodology for equivalent checking providing the foundation for corresponding solutions in the (near) future.
\end{abstract}

\section{Introduction}
\label{sec:Introduction}
Since the introduction of the first transistor, the semiconductor industry has made exponential improvements for decades. Accordingly, the resulting circuits and systems substantially increased in complexity---constantly making it harder to properly verify their correctness. Equivalence checking, i.e., verifying whether two circuits realize the same functionality or not, is a typical task in this context. But despite the fact that, in the meantime, numerous approaches addressing this task have been proposed (see, e.g., \cite{1097859,furter_1,bryant_1986,armin1999symbolic, further_sat_1, further_sat_2, disch_2013,580110}), all approaches still suffer from the steadily increasing complexity and the resulting verification gap (i.e., the fact that the designs grow faster than the ability to efficiently verify them).

Consequently, the verification community should consider all available alternative directions and explore whether they may provide additional support to eventually close or at least narrow this gap. Quantum computing~\cite{NC:2000} could provide such an alternative. Although still in its infancy, recent achievements have already sparked interest in many application domains, such as finance~\cite{stamatopoulosOptionPricingUsing2020}, chemistry~\cite{peruzzoVariationalEigenvalueSolver2014}, machine learning~\cite{zoufalQuantumGenerativeAdversarial2019}, and optimization~\cite{harwoodFormulatingSolvingRouting2021}.
Even if the resulting solutions are often not practical yet and still restricted by the limited hardware capabilities, these considerations already indicated the potential for quantum advantages in the (near) feature. 
This raises the question whether similar quantum advantages can be achieved for equivalence checking as well.

In this work, we are aiming to provide a first step towards answering this question. 
While we are not claiming yet that quantum computing certainly will provide an advantage for equivalence checking, we propose a conceptual basis to tackle this task with this new computational concept. 
To this end, we are considering Grover's algorithm~\cite{grover1996fast}---a quantum solution for which quantum advantage already has been verified on a conceptual level. 
We show that this seems to be an obvious choice for equivalence checking as well, but also illustrate some pitfalls that might lead to an improper application of Grover and, hence, the generation of meaningless results. 
Afterwards, we discuss how these pitfalls can be addressed---resulting in a concept and workflow for equivalence checking using quantum computing.

Experimental evaluations confirm the applicability of the proposed approach and show that, even despite the probability nature of quantum computing, still helpful conclusions can be drawn from the obtained results---providing verification engineers with a \enquote{starting point} to an alternative computing technology that may help them to solve the equivalence checking problem.

The remainder of this work is structured as follows.
In \autoref{sec:rel_work}, we discuss the basics of 
Grover's algorithm.
\autoref{sec:Concepts} presents a \mbox{straightforward} approach to equivalence checking using quantum computing and shows its pitfalls, which can easily lead to meaningless results. 
How these can be addressed and how a workflow can be derived accordingly is then covered in \autoref{sec:Handling Grover}. 
The resulting workflow is then evaluated and discussed in \autoref{sec:Evaluation} before \autoref{sec:Conclusion} concludes this work.

\section{Background}
\label{sec:rel_work}
To keep this work self-contained, this section briefly reviews one of the most prominent quantum computing algorithms: the \emph{Grover's} algorithm.

\subsection{Grover's Algorithm}\label{sec:Grover}
\emph{Grover's algorithm}~\cite{grover1996fast} provides a quadratic \mbox{speed-up} for several problems compared to classical algorithms. 
This algorithm was originally motivated by the problem of finding desired elements $\hat{x}$ in an unordered database, but it can be applied to further search problems as well, e.g., to determine the desired basis states out of all $2^n$ possibilities for $n$ qubits. 
Formally, the problem is defined by having a function $f:\mathbb{B}^n \longrightarrow \mathbb{B}$ and finding an $\hat{x}\in\mathbb{B}^n$ such that $f(\hat{x})=1$.

\vspace{1cm}

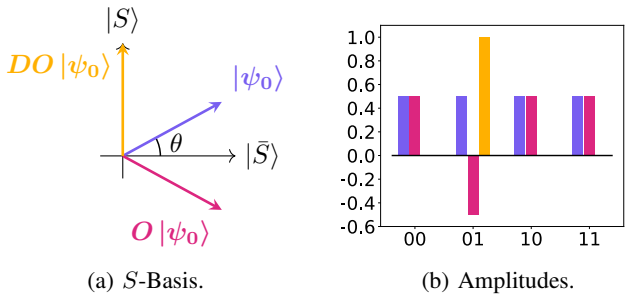
\begin{figure}[t]
    \centering
    \begin{subfigure}[b]{0.23\textwidth}
        \centering
        \begin{tikzpicture}[scale=1.5]
        \coordinate (A) at (0,0);
        \coordinate (B) at (1,0);
        \coordinate (C) at (0.878,0.479);
        \pic [draw, -,"$\theta$",line width = 0.2mm, pic text options={shift={(12pt,3pt)}}] {angle = B--A--C};
        \draw[->] (-0.2,0)--(1,0) node[right]{$\ket{\Bar{S}}$};
        \draw[->] (0,-0.2)--(0,1) node[above]{$\ket{S}$};
        \draw[line width=1pt,MyPurple,-stealth](0,0)--(0.878,0.479) node[anchor=south west]{$\boldsymbol{\ket{\psi_0}}$};
        \draw[line width=1pt,MyPink,-stealth](0,0)--(0.878,-0.479) node[anchor=north east]{$\boldsymbol{O\ket{\psi_0}}$};
        \draw[line width=1pt,MyYellow,-stealth](0,0)--(0,1) node[anchor=north east]{$\boldsymbol{DO\ket{\psi_0}}$};
        
        \end{tikzpicture}
        \caption{$S$-Basis.}
        \label{fig:Grover_state vector}
    \end{subfigure}
    \hfill
    \begin{subfigure}[b]{0.23\textwidth}
        \centering
        \includegraphics[width=\textwidth]{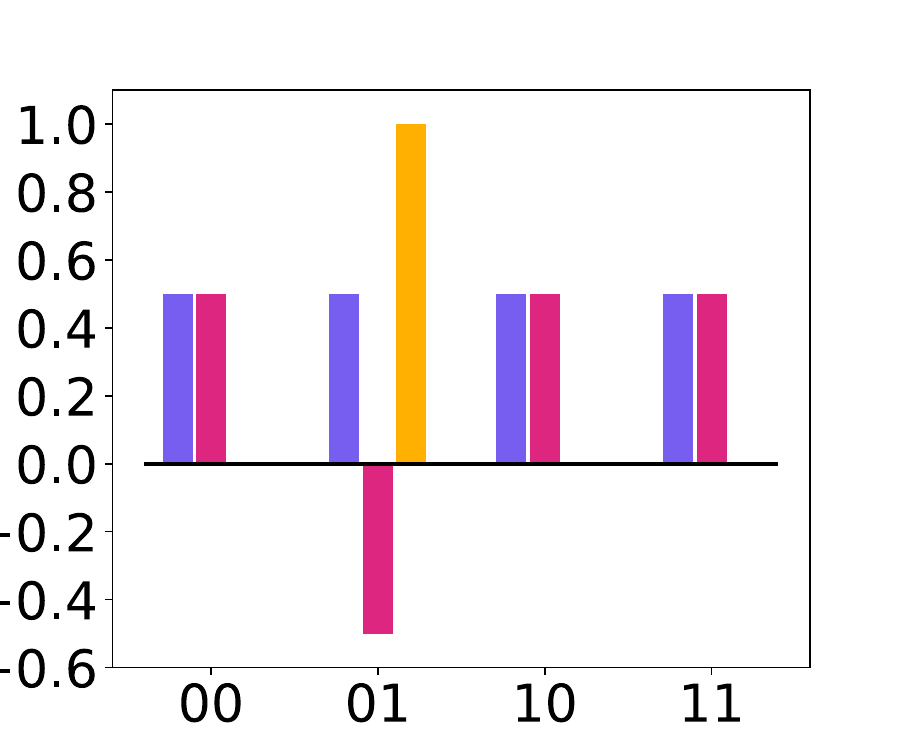}
        \caption{Amplitudes.}
        \label{fig:grover_amplitudes}
    \end{subfigure}
    \caption{State vector evolution in the $S$-Basis with the corresponding state amplitudes.}
    \label{fig:Grover}
 \vspace{-5mm}
\end{figure}

The algorithm itself consists of three steps (state preparation, oracle operator, and diffusion operator) which are reviewed in the following. To make the algorithm more accessible, the original input and the results of each step are visualized on a plane shown in \autoref{fig:Grover_state vector} that is spanned between the superposition of all desired states $\ket{S}$ and all the other remaining states $\ket{\bar{S}}$ that are orthogonal to each other. 
In the following, we will refer to this as the \emph{S-Basis}. 
The respective amplitudes of the states are shown in \autoref{fig:grover_amplitudes} in matching colors. 

\begin{exmp}\label{ex:s_not_s}
    Let us assume a database over 2 bits and, hence, 
    four possible inputs $00$, $01$, $10$, $11$, where \mbox{wlog.~$\hat{x}=01$} represents the desired state. Thus, $\ket{S} = [0~1~0~0]^T$ and $\ket{\bar{S}} = 1/\sqrt{3}[1~0~1~1]^T$ form the coordinate system (as shown in \autoref{fig:Grover_state vector}).
    Obviously, these states are not known beforehand, but are used to visualize the results of the algorithm.
\end{exmp}

\emph{Step 1 (State Preparation)}: Initially, all qubits are put in a \emph{superposition} such that the amplitudes of all basis states are equally distributed. This can be easily done by starting with a basis state~$\ket{00}$ and applying H operations to all qubits.

\begin{exmp}\label{ex:h_state}
    The result of the first step is shown by the pink vector in \autoref{fig:Grover_state vector} and the respectively colored amplitude in \autoref{fig:grover_amplitudes}. 
    This state is expressed in terms of the \emph{S-Basis} by 
    $\ket{\psi_0} = H^{\bigotimes2}\ket{0} = \frac{1}{2}\ket{S} + \frac{\sqrt{3}}{2}\ket{\bar{S}} = \frac{1}{2} [1~1~1~1]^T.$
\end{exmp}
\emph{Step 2 (Oracle Operator)}: This step is the most important one and \emph{encodes} the actual problem, i.e., it is also the only step that is problem-dependent.
To this end, the function~$f$ (representing the problem) is implemented as an \emph{oracle}~$O$ which \emph{marks} the amplitudes for all basis states representing the desired input allocations~$\hat{x}$ with a %
negative phase, while keeping all other amplitudes (representing inputs~$x\neq\hat{x}$) unchanged.
That is, the application of the oracle effectively mirrors the initial state $\ket{\psi_0}$ around $\ket{\bar{S}}$. 
Therefore, it can be written as $O = I - 2\ket{S}\bra{S}$ with $I$ being the identity operator.
Generating quantum circuits that realize such oracles has been actively researched for years and many \mbox{so-called} \emph{reversible logic synthesis} tools have been proposed for this purpose, such as~\cite{seidel2023automatic, willeBDDbasedSynthesisReversible2009, zilic2007reversible, sanaee2010esop, zulehnerOnepassDesignReversible2018, date2017_3}.

\begin{exmp}\label{ex:oracle_application}
Recall that wlog., $\hat{x}=01$ represents the desired input.
Then, the result of the second step (i.e., the result of applying the oracle~$O$ to the state obtained in \autoref{ex:h_state}) is shown by the pink vector in \autoref{fig:Grover_state vector} and its respective amplitude in \autoref{fig:grover_amplitudes}.
  This state is expressed in terms of the \emph{S-Basis} by 
    $\ket{\psi_1} = -\frac{1}{2}\ket{S} + \frac{\sqrt{3}}{2}\ket{\bar{S}}$.
It can be clearly seen that this state now explicitly marks the basis state representing the desired input~$\hat{x}$, i.e., the corresponding amplitude of the state $01$ is negative, while all other amplitudes (representing basis states $x\neq\hat{x}$) remain unchanged.
\end{exmp}

Now, one may assume that the problem is solved, since the resulting state clearly marks the desired input. 
But the state cannot be directly observed and must be measured. 
However, a measurement of this state still yields all basis states with the same probability (in the case of the example, both~$|\frac{1}{2}|^2$ and $|-\frac{1}{2}|^2$ yield 0.25, i.e., all basis states are still measured with the same probability).
Therefore, a third and last step is necessary.

\emph{Step 3 (Diffusion Operator):} This step conducts a \mbox{so-called} \emph{amplitude amplification}, i.e., it aims to isolate the desired states by increasing the amplitudes (and, hence, the probability of measurement) of the basis states marked by a negative phase, while decreasing the amplitudes (and, hence, the probability of measurement) of all other basis states. 
This is done by applying $D = 2 \ket{\psi_0}\bra{\psi_0} - I$, where $\ket{\psi_0}$ is the initial state.
Applying this operator effectively mirrors the state $O \ket{\psi_0}$ around $\ket{\psi_0}$.

\begin{exmp}\label{ex:diffusion_application}
    
Applying the diffusion operator to the result of the second step shown in \autoref{ex:oracle_application} rotates the corresponding vector by $4\theta=4\arcsin{\sqrt{\frac{1}{4}}} = \frac{2\pi}{3}$. This eventually yields the final result, which is shown by the yellow vector and amplitude in \autoref{fig:Grover}.
One can clearly see, that measuring this state now yields the desired basis state~$\hat{x}$ with a substantially higher probability (in fact, here even with $100\%$ probability, since $|1|^2=1$) while all other basis states are measured with a substantially lower probability (in fact, here even with $0\%$ probability, since $|0|^2=0$). 
Considering that the actual function~$f$ was only evaluated once (when applying the oracle) and not $2^n$ times, this constitutes a promising advantage over classical approaches. 

\end{exmp}

In order to make that work, it is imperative that particularly Steps 2 and~3 are \emph{properly} applied.
They form a \mbox{so-called} \emph{Grover iteration} and may need to be applied more than once to actually get the best probabilities. 

However, the optimal number of iterations actually depends on the number of desired states and the number of total states.
Selecting a \mbox{non-ideal} number of iterations could negatively affect the desired probabilities and, in the worst case, lead to random measurement results.
Furthermore, note that the application of these steps does not always guarantee a $100\%$ probability of finding the desired element despite an optimal number of iterations.

\subsection{Probability of Measuring Target States}\label{sec:Grover_Probability}

The probability of measuring the desired states (and, hence, their distinctiveness from the remaining states) after a certain number of Grover iterations depends on the angle between $\ket{\psi_0}$ and $\ket{\Bar{S}}$ and, therefore, on the number of desired states $c$ and total states $N$. This angle $\theta$ is given by

\[
\theta = \arcsin{(\sqrt{\frac{c}{N}})}.
\]

Since every Grover iteration mirrors the current state vector first around $\ket{\Bar{S}}$ and, then, around $\ket{\psi_0}$, the angle $\theta$ first evolves to $-\theta$ and, then, to $3\theta$. 
The resulting vector contains a component parallel to $\ket{S}$ of $\sin{((2g+1)\theta)}$ after $g$ iterations. 
Since the probability of measuring a state is the square of its amplitude, the probability of measuring one of the desired states, which are the basis states contained in $\ket{S}$, after $g$ iterations is given by
\[
    P = \sin^2((2g+1)\theta) = \sin^2((2g+1)\arcsin{(\sqrt{c/N})}).
\]

\begin{figure}[t]
    \centering
    \includegraphics[scale=0.22]{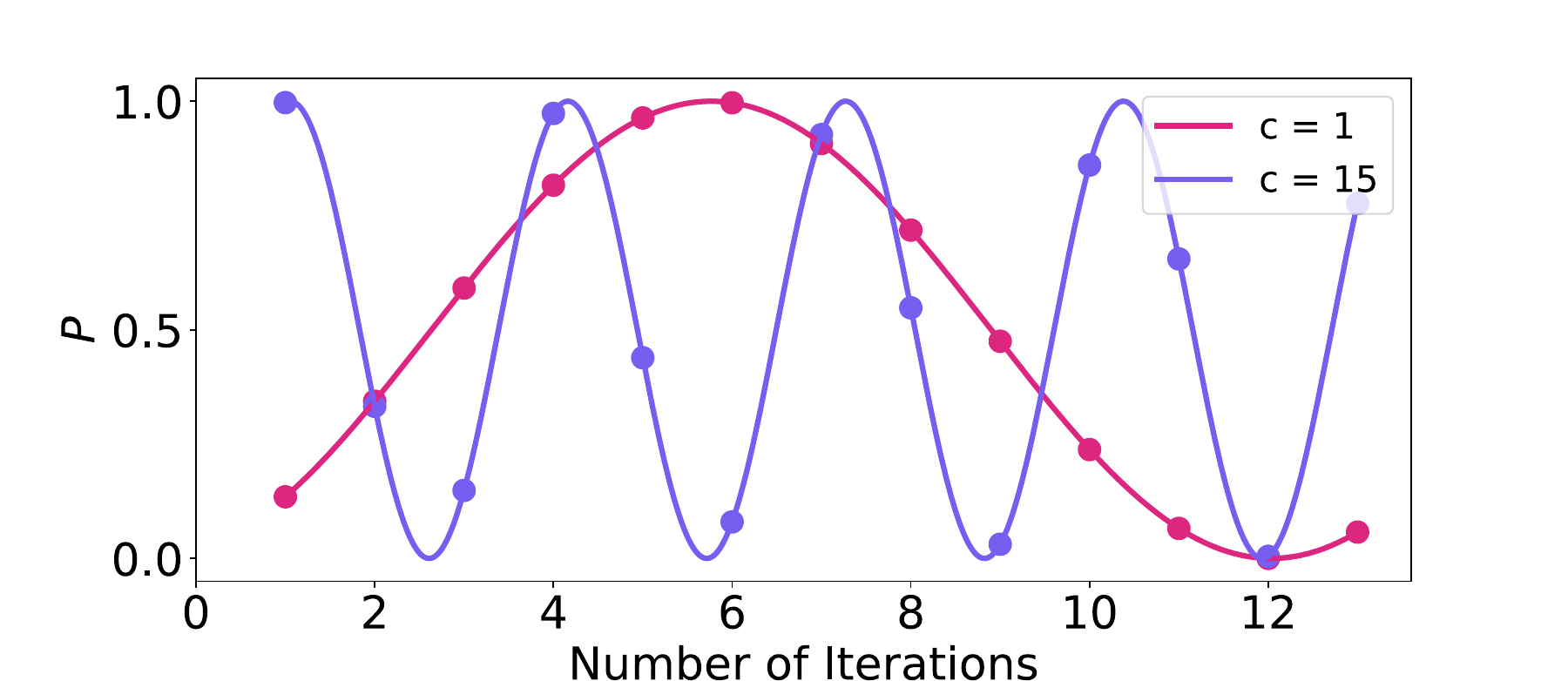}
    \caption{Probability of measuring a desired state in dependence of the number of iterations when $1$ respectively $15$ out of $64$ basis states represent desired states.}
    \label{fig:Grover_probabilities}
 \vspace{-5mm}
\end{figure}

Hereby, the probability of measuring any basis state \emph{within} one of both two groups of states---the desired or remaining ones---is always evenly distributed. 
However, not every probability of measuring desired elements is achievable because the number of Grover iterations is an integer.

Furthermore, if $P$ goes to zero, every state except the desired ones will be measured, which also allows a distinction of both groups.

\begin{exmp}
    The probability of measuring a desired state is denoted in \autoref{fig:Grover_probabilities} for two examples with different numbers of desired states. 
    In the case of $1$ desired state, the sinus function has a much lower frequency compared to the $15$ desired states case---highlighting the importance of choosing a suitable number of Grover iterations.
    This becomes even clearer when considering the probabilities for the $g=5$ Grover iterations:
    While for the $1$ desired state case the probability is almost $100\%$, it is close to $50\%$ in the other case.
\end{exmp}

\section{Equivalence Checking\\Using Quantum Computing}\label{sec:Concepts}

This section briefly reviews classical equivalence checking and motivates how this task could be tackled using quantum computing.

\subsection{Classical Equivalence Checking}\label{sec: Classical Equivalence Checking}
Designing classical circuits and systems is a tedious process that requires multiple levels of abstraction, as well as several \mbox{non-trivial} steps to implement a given functionality in silicon.
Starting from a mathematical definition, the specification is synthesized using the actually available hardware gates and potentially optimized until it is finally realized on a physical computer chip.
Through all those steps, it is of utmost importance to ensure that the initial specification is always met and the functionality is never altered---using equivalence checking.

The problem of equivalence checking is simple to formulate: Given two circuits designed to realize the same target, it shall be proven whether they realize the same functionality or not. In the latter case, the non-equivalence shall be witnessed by a counter example, i.e., an input pattern which, when applied to both circuits, yields different output values. 
Over the past decades, numerous approaches have been proposed to tackle that problem~\cite{1097859,furter_1,bryant_1986,armin1999symbolic, further_sat_1, further_sat_2, disch_2013,580110}---methods based on
Binary Decision Diagrams~(BDDs, \cite{bryant_1986}) 
or \emph{Boolean satisfiability}~(SAT, \cite{armin1999symbolic, disch_2013}) are two prominent examples.
Furthermore, also \enquote{unusual} approaches are frequently evaluated (see, e.g.,~\cite{AGWD:2016,aspdac_2023_3}).

In this work, we consider the problem through the \emph{miter structure} as described in~\cite{580110}.
Here, two circuits to be checked are applied with the same primary inputs. Then, for each pair of \mbox{to-be-equal} output bits, an \mbox{exclusive-OR} (XOR) gate is applied---evaluating to~$1$ if the two outputs generate different values (which only happens in the case of \mbox{non-equivalence}).
By \mbox{OR-ing} the outputs of all these XOR gates, eventually an indicator results that shows whether both circuits are equivalent. Then, the goal is to determine an input assignment so that this indicator evaluates to~1 (providing a counter example that shows non-equivalence) or to prove that no such assignment exists (proving equivalence).

\begin{figure}[t]

	\centering 
 \resizebox{0.64\linewidth}{!}{
\begin{circuitikz} \draw
(0,1.2) node[and port] (and1) {\footnotesize{AND}}
(0,0) node[xor port, number inputs=3] (xor1) {\footnotesize{XOR}}
(1.8,0.5) node[and port] (and2) {\footnotesize{AND}}
(and1.out) -- (and2.in 1)
(xor1.out) -- (and2.in 2)

(0,-1.5) node[and port] (and3) {\footnotesize{AND}}
(0,-2.7) node[or port, number inputs=3] (or1) {\footnotesize{OR}}
(1.8,-2.0) node[and port] (and4) {\footnotesize{AND}}
(and3.out) -- (and4.in 1)
(or1.out) -- (and4.in 2)

(3.5,0.22) node[xor port] (xor2) {\footnotesize{XOR}}
(and4.out) |- (xor2.in 2)
(and2.out) |- (xor2.in 1)
(3.5,-1.0) node[xor port] (xor3) {\footnotesize{XOR}}
(and1.out) |- (xor3.in 1)
(and3.out) |- (xor3.in 2)
(5.2,-0.35) node[or port] (or2) {\footnotesize{OR}}
(xor2.out) -- (or2.in 1)
(xor3.out) -- (or2.in 2)

(and1.in 1) node [anchor=east] {$x_0$}
(and1.in 2) node [anchor=east] {$x_1$}
(xor1.in 1) node [anchor=east] {$x_2$}
(xor1.in 2) node [anchor=east] {$x_3$}
(xor1.in 3) node [anchor=east] {$x_4$}

(and3.in 1) node [anchor=east] {$x_0$}
(and3.in 2) node [anchor=east] {$x_1$}
(or1.in 1) node [anchor=east] {$x_2$}
(or1.in 2) node [anchor=east] {$x_3$}
(or1.in 3) node [anchor=east] {$x_4$};

 \node[fit=(and1) (and2) (xor1), draw, blue, dashed, inner sep=1.5pt, label={Circuit A}] (rect) {};
  \node[fit=(and3) (and4) (or1), draw, blue, dashed, inner sep=1.5pt] (rect2) {};
  \node[below=of rect2, yshift=10mm] {Circuit B};
    \node[fit=(xor2) (xor3) (or2), draw, red, dashed, inner sep=1.5pt, label={Miter}] (rect3) {};
\end{circuitikz}}
	\caption{Miter configuration with two non-equivalent circuits.}
	\label{fig:miter}
 \vspace{-5mm}
\end{figure}
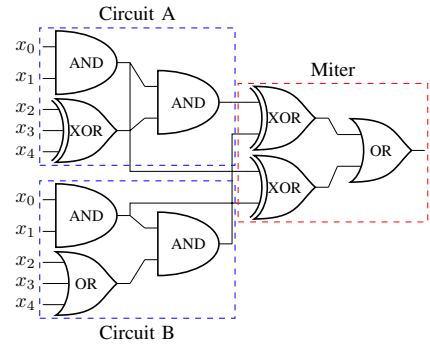

\begin{exmp}\label{ex:miter}
\autoref{fig:miter} shows two circuits A and B, each with five inputs and two outputs.
    Although they are obviously different since they comprise different gates---Circuit A comprises an XOR gate, while Circuit B comprises an OR gate---it is unclear whether they nevertheless realize the same functionality.
    For that, the miter structure is applied and the two output bits of both circuits are connected with two XOR and one OR gate. 
    Then, to check for equivalence, an input assignment must be determined so that the indicator evaluates to~$1$ (or to prove that such assignment does not exist). Here, this is the case, e.g., for the input assignment~$11111$.
\end{exmp}

\subsection{Using Grover's Algorithm for Equivalence Checking}

At first glance, the equivalence checking problem reviewed above seems very similar to the database search problem reviewed in \autoref{sec:rel_work}.
In both cases, a function~$f$ exists (a database or the miter-formulation) for which an~$\hat{x}$ shall be determined so that~$f$ evaluates to $f(\hat{x})=1$.
Accordingly, using Grover's algorithm for this problem could be considered a good first approach.
To this end, only the oracle (representing the actual problem; thus far: database search, now: equivalence checking) needs to be replaced.
This can be done using a reversible circuit synthesis method (e.g., as proposed in~\cite{seidel2023automatic, willeBDDbasedSynthesisReversible2009, zilic2007reversible, sanaee2010esop, zulehnerOnepassDesignReversible2018, date2017_3}).

If \enquote{properly} executed, a Grover approach should generate a quantum state in which all desired basis states representing counter examples have a much larger amplitude (and, hence, probability of getting measured) than the remaining states. 
Since a separation of these groups into as distinct as possible subsets is desired, a \enquote{proper} number of Grover iterations should be chosen.
However, determining an optimal number of iterations requires information on the number of elements that one is looking for.
Since it is part of the actual problem to find out if and how many such elements exist, this information is not accessible.

Therefore, both the equivalence and the non-equivalence cases can be boiled down to choosing a suitable number of Groover iterations and postprocessing the resulting measurements. 
For the former, every number of iterations leads to the same superposition outcome and, therefore, a procedure must be defined to distinguish the desired outcome from cases in which there are actual counter examples, but an unfortunate number of Grover iterations has mistakenly led to a superposition. 
For the latter, the number of Grover iterations leading to the most distinguishable separation of the counter examples from the remaining states must be determined.

\section{Handling the \\ unknown Number of Grover Iterations}\label{sec:Handling Grover}
This section describes the proposed approach on how to apply Grover’s algorithm to the equivalence checking problem without knowing the number of counter examples. 
To this end, we first cover conceptually how the proposed approach can detect non-equivalence and equivalence. Afterwards, we provide a description of a possible implementation of this concept.

\begin{figure}[t]
    \centering
    \includegraphics[scale=0.5]{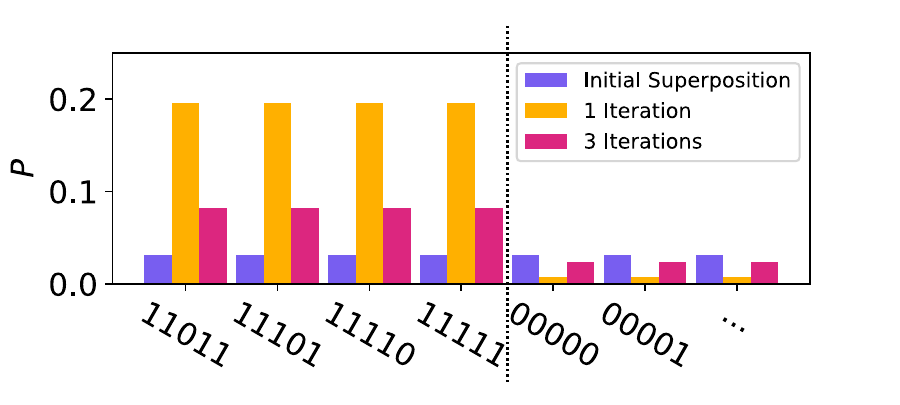}
    \caption{Probability distributions of 32 states for different numbers of iterations with the four counter examples "11011", "11101", "11110", and "11111".}
    \label{fig:state vector-histogramm}
 \vspace{-5mm}
\end{figure}

\subsection{Concept}
In the case of \emph{non-equivalence}, Grover's algorithm isolates the basis states that represent counter examples from the remaining ones by obtaining different probabilities for the two groups of being measured. 

For that, a suitable number of Grover iterations must be chosen and, for that, almost every number of iterations is sufficient to isolate the basis states of counter examples from the ones of the remaining states.
Although different numbers lead to different probabilities of measuring the counter examples' basis states, they are clearly distinguishable for \emph{any} number of iterations---except for two corner cases, which are described later.
Hereby, it is irrelevant which of the two groups of basis states has a larger probability of being measured, since by simulating the circuits with one of the basis states as input, the group containing the counter examples can be determined.
Nevertheless, it is desired to derive a number of Grover iterations that allows us to achieve the most distinct isolation of the basis states of counter examples from the remaining ones.

In the case of \emph{equivalence}, it is different, since \emph{any} number of Grover iterations will result in an equal superposition of \emph{all} basis states.
However, since \emph{any} number of iterations should lead to a difference in the probabilities of basis states in the case of non-equivalence, the argumentation can be turned around: If none of the evaluated numbers of Grover iterations has led to a distinction between basis states, it can be deduced that the considered equivalence checking instance is actually equivalent.

\begin{exmp}\label{ex:hist-state vector}
    \autoref{fig:state vector-histogramm} illustrates this approach for the case shown in \autoref{fig:miter} with four counter examples out of 32 basis states. 
    Applying either one or three Grover iterations isolates the desired states sufficiently enough to distinguish them from the remaining ones (indicated by the dashed line in \autoref{fig:state vector-histogramm}). 
    However, the isolation is significantly more distinct for one iteration compared to three. 
    If both circuits were equivalent, the distributions would have led to an equal superposition regardless of the number of iterations as illustrated in purple.

\end{exmp}

Nevertheless, for some combinations of the number of counter examples $c$ and the total number of states $N$, the two groups of basis states cannot be separated---leading to the two previously mentioned corner cases in which no isolation of the basis states of the counter examples from the ones of the remaining states is possible.

First, if $\frac{c}{N}=0.5$, the angle $\theta$ of the state vector in \autoref{fig:Grover_state vector} alternates between 45\degree and 135\degree. For both angles, an equal superposition of all basis states is obtained.
Second, if the sum of probabilities of measuring the basis states representing counter examples is equal to $\frac{c}{N}$ (due to an unfortunately chosen number of iterations), a superposition is generated as well.

\subsection{Implementation}\label{sec:Realization}

Direct access and evaluation of the probabilities of the basis states as described above only works conceptually or when simulators such as~\cite{zulehnerAdvancedSimulationQuantum2019, sim1,hillmichJustRealThing2020, vincentJetFastQuantum2021, villalongaFlexibleHighperformanceSimulator2019, date2019_1} are used to determine the corresponding amplitudes of a state.

\begin{figure}[t]
    \centering
    \includegraphics[scale=0.5]{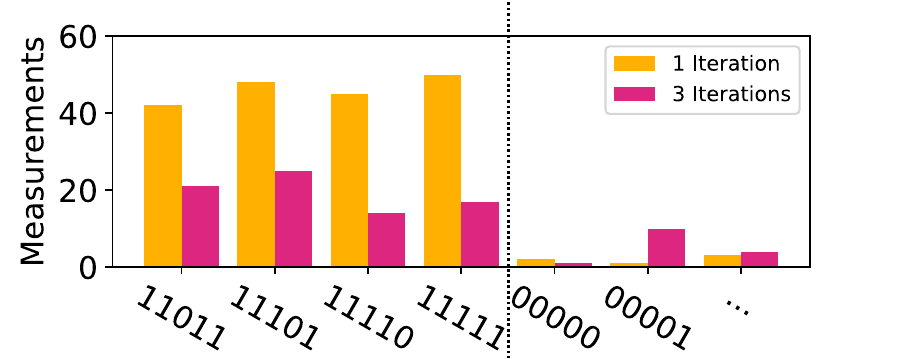}
    \caption{Distribution of number of measurements when executed with 256 shots.}
    \label{fig:histogramm-shot-based}
 \vspace{-5mm}
\end{figure}

When executing on a real device, only the measurement outcome is accessible instead of the actual probabilities of the measurement outcome.
To approximate those probabilities as closely as possible, a given quantum circuit is usually executed several times---the so-called \emph{shots}.

Due to this approximation, the proposed approach must be extended. 
Although, on a conceptual level \emph{any} number of Grover iterations could be used since all numbers lead to a difference between the two categories of basis states---unfortunate numbers lead to a less distinct one.
If the difference is now too small, the approximation effect could compromise the result when its induced inaccuracy is larger.

\begin{exmp}
    When evaluating the same equivalence checking instance as in \autoref{ex:hist-state vector} with 256 shots instead of directly accessing the basis state probabilities, exemplary results for one and three Grover iterations are shown in \autoref{fig:histogramm-shot-based}.
    In contrast to before, the result after three applied iterations is not as distinct to evaluate, as the counter example state "11110" was measured only four times more often than "00001". 
    Applying one Grover iteration yields a clearer separation of the basis states representing counter examples from the remaining ones (which is, again, indicated by the dashed line in \autoref{fig:state vector-histogramm}).   
\end{exmp}

Therefore, a number of Grover iterations leading to an isolation of the counter examples' basis states from the remaining ones as distinct as possible is required. 
For this purpose, an approach based on \cite{zalka1999grover} is designed that tests different numbers of iterations until the isolation of a group of basis states is reached. 
It starts with the number of Grover iterations for the case of one counter example and verifies if the measurements of two groups of basis states are sufficiently distinct.

To decide this, the basis states are sorted with respect to how often they are measured from largest to smallest. 
Starting with the most frequently measured basis state, it is compared to the subsequent state.
To determine whether their difference is large enough to mark the division between the two groups of states, an adjustable parameter $\phi$ is used to define a threshold, which is reached if 
\[
    1 - \frac{m_{i+1}}{m_{i}} > \phi
\]
\noindent with $m_i$ being the number of measurements of the basis state with index $i$. 

This procedure (which is shown in \autoref{alg:algorithm}) is repeated until either the equation validates to true---dividing all basis states into the two groups representing counter examples and the remaining states---or all measured states (or at most $\frac{N}{2}$ states) have been iterated. 
The latter case could indicate two things: Either \emph{all} measured states are counter examples or the combination of the used $\phi$ value and Grover iteration number was not suitable.
This can be differentiated by comparing the number of measured states $|m|$.
If $|m|\leq N/2$, all measured states are counter examples, because one of the two separated groups of states must be smaller or equal to $N/2$, but the threshold was not exceeded. 
Therefore, the probability of measuring one group of basis states was so large that no other states were measured. 
If $|m|>N/2$ is greater, the number of Grover iterations is reduced by one.
If one group of states is found to be sufficiently distinct, the circuits must be simulated with one of the basis states as input to determine which of the two groups represents the counter examples.
If none of the tested numbers of Grover iterations yields a sufficient distinction between two groups of basis states, the circuits are assumed to be equivalent, as no counter examples could be found.

\begin{algorithm}[t]
	\caption{Implementation} 
        \label{alg:algorithm}
	\begin{algorithmic}[1]
            \State $g\leftarrow$ number of Grover iterations assuming $c=1$
            \State $\phi\leftarrow$ parameter value
		\While {$g>0$}
                \State run Grover's algorithm and sort measurement outcomes $m$ decreasingly
                \State $|m|\leftarrow$ number of measured basis states
            
			\For {$i=1,2,\ldots,\frac{N}{2}$}
                    \If {$1-\frac{m_{i+1}}{m_i}>\phi$ or $i==|m|$}
                    \State determine counter examples 
				\State return non-equivalence with counter examples 
                    \EndIf
			\EndFor
                \State $g\leftarrow g-1$ 
            \EndWhile
            \State return equivalence
	\end{algorithmic} 
\end{algorithm}

This approach enables the solving of equivalence checking problems with quantum computing for arbitrary instances---if equivalent or not and with how many counter examples for the latter.

\section{Experimental Evaluation}
\label{sec:Evaluation}

The approach described in \autoref{sec:Realization} has been evaluated with equivalence checking instances from $6$ to $9$ bits comprising a different number of counter examples~$c$.
To this end, \autoref{alg:algorithm} has been implemented in Python on top of Qiskit $0.42.0$\footnote{The proposed method has been integrated into the MQT~ProblemSolver that is part of the Munich Quantum Toolkit (MQT;~\cite{willeMQTHandbookSummary2024}) and publicly available at \url{https://github.com/cda-tum/mqt-problemsolver}.}. 
For the quantum circuit execution itself, a simulator has been used, and each problem instance is run $10$ times (each executed with $8\cdot N$ shots and $N$ being the total number of possible states) to accommodate the probabilistic nature of this quantum computing approach.
In this section, we present the obtained results and, afterwards, discuss the correspondingly obtained insights towards a working concept of a quantum computing methodology for equivalence checking. 

\subsection{Results}
As described above, the objective is to reach a state in which either of the two groups of basis states have a sufficiently distinct probability (in case the circuits are non-equivalent) or all basis states have the same probability (in case the circuits are equivalent).
Although conceptually \emph{any} difference between the amplitudes of two basis states indicates that one is a counter example and the other one is not, this difference is blurred by the number of shots that approximate the underlying basis states' amplitudes.
Therefore, in the proposed approach, a threshold parameter $\phi$ is used to define how distinct the measurement outcomes of two basis states must be to be \emph{sufficiently distinct} (cf. \autoref{sec:Realization}).
Obviously, the value of $\phi$ has a substantial impact on the applicability of the proposed approach. 
Consequently, we exhaustively evaluated its influence. 
To this end, all instances (including equivalent circuits as well as circuits with different numbers of counter examples) have been checked with the proposed approach---using values for $\phi$ from $0.1$ to $0.9$. 

The results are summarized in \autoref{tab:results}. 
Here, the columns indicate the number of inputs, the number of counter examples ($0$~indicates that the circuits are equivalent), as well as the different values of $\phi$.
The table entries \enquote{-} indicate that the circuits have been classified falsely (classifying \mbox{non-equivalent} circuits as equivalent or vice versa), while the numbers indicate a correct classification and express the necessary accumulated number of Grover iterations (thus, lower values are desirable). 

The results clearly show the influence of the threshold parameter $\phi$: 
While for a small value of $\phi=0.1$ only a few instances could be verified, this changes with an increasing value.
For $\phi$ between $0.3$ and $0.7$, all problem instances were correctly verified.
However, there is a turning point:
When~$\phi$ becomes too large, the \mbox{non-equivalent} instances are falsely classified since the threshold cannot properly distinguish the probabilities of basis states that represent counter examples from the remaining ones anymore.
In addition to that, a larger value for $\phi$ potentially also leads to a larger number of required Grover iterations.

Hence, $\phi$ offers a trade-off between the reliability of the results and the required number of Grover iterations.
In general, $\phi$ should be as small as possible but large enough to detect actual differences in the measurement outcomes.
With this knowledge, we can now formulate a working concept of how equivalence checking problems can potentially be solved using quantum computing in the future.

\begin{table}[t]
    \centering
    \caption{Obtained results.}

    \begin{tabular}{c|c|ccccc}
    Input  & Counter  & \multicolumn{5}{c}{$\phi$} \\ 
     Bits& Examples& 0.1  & 0.3 & 0.5 & 0.7 & 0.9 \\ \hline
        6 & 0 & - & 15 & 15 & 15 & 15  \\ 
        6 & 1 & 5 & 5 & 5 & 5 & 5  \\ 
        6 & 3 & - & 5 & 5 & 5 & 9  \\ 
        6 & 6 & - & 12 & 12 & 12 & 14  \\ 
        6 & 13 & - & 5 & 5 & 5 & -  \\ \hline
        7 & 0 & - & 36 & 36 & 36 & 36  \\ 
        7 & 1 & 8 & 8 & 8 & 8 & 8  \\ 
        7 & 6 & - & 8 & 8 & 17 & 30  \\ 
        7 & 13 & - & 8 & 9 & 15 & 15  \\ 
        7 & 26 & - & 8 & 8 & 8 & 8  \\ \hline
        8 & 0 & - & 78 & 78 & 78 & 78  \\ 
        8 & 3 & - & 12 & 12 & 12 & 23  \\ 
        8 & 13 & - & 12 & 12 & 22 & 23  \\ 
        8 & 26 & - & 12 & 12 & 12 & 12  \\ 
        8 & 51 & - & 12 & 20 & 23 & 50  \\ \hline
        9 & 0 & - & 153 & 153 & 153 & 153  \\ 
        9 & 5 & - & 17 & 17 & 17 & 80  \\ 
        9 & 26 & - & 17 & 17 & 17 & 17  \\ 
        9 & 51 & - & 17 & 17 & 17 & 17  \\ 
        9 & 102 & - & 48 & 48 & 48 & - \\  %
    \end{tabular} %
    \label{tab:results}
 \vspace{-5mm}
\end{table}

\subsection{Resulting Working Concept}

Based on all the discussions from above, a potential future workflow for equivalence checking using quantum computing would basically be composed of two main steps.

First, the corresponding equivalence checking instance needs to be translated into a quantum computing formalism.
That is, the problem needs to be encoded and, then, inserted as an oracle into Grover's algorithm.
To this end, existing tools for reversible circuit synthesis (such as, e.g.,~\cite{seidel2023automatic, willeBDDbasedSynthesisReversible2009, zilic2007reversible, sanaee2010esop, zulehnerOnepassDesignReversible2018, date2017_3}) can be utilized. 
Although the paradigms are substantially different, from a verification engineer's perspective, this is quite similar to the conventional equivalence checking flow (where the problem instance also needs to be, e.g., represented as decision diagram~\cite{bryant_1986}) or encoded as SAT instance~\cite{armin1999symbolic, disch_2013}.

Afterwards, the resulting quantum circuit has to be executed following \autoref{alg:algorithm} and the obtained results have to be interpreted. 
Due to the probabilistic nature of quantum computing, this is very different now, since, e.g., a number of measurements is used to approximate the underlying quantum state.
In the evaluations summarized above, $8 \cdot N$ shots were sufficient for this. 

However, as discussed, using a proper $\phi$ greatly affects the probability of determining the correct result.
The value should be as small as possible because larger values result in more Grover iterations and, thus, computational effort. 
The above evaluations indicate that $\phi=0.3$ is promising.
In conclusion, finding no counter examples with this setting is a strong indication that both circuits are equivalent.
Otherwise (if both circuits are not equivalent), the proposed approach delivers counter examples with a high probability, which then can simply be confirmed by simulating the circuits with the corresponding inputs.

To further improve the proposed realization, the number of shots used can be increased.
This results in a measurement outcome distribution that better mimics the underlying quantum states---allowing for a smaller $\phi$ value that causes less computational effort since fewer Grover iterations are executed. 

Overall, the approach provides the verification engineer with a working concept to carry out equivalence checking with quantum computing.
However, its performance is influenced by two factors: the number of used shots and the value of the parameter~$\phi$. 
While explicitly evaluating these influences and deriving associated probabilities for the correctness of the determined result---either empirically or theoretically---remains for future work, the resulting concept provides a promising alternative to potentially tackle the verification gap in the future using quantum computing.

\section{Conclusion}\label{sec:Conclusion}
Equivalence checking of classical circuits is an important task in the semiconductor industry.
Although various solutions have been proposed in recent decades to address the steadily increasing complexity, there still exists a verification gap, since the circuit design grows faster than the ability to efficiently verify it.
In this work, a different approach based on quantum computers has been explored.
To this end, the equivalence checking problem has been formulated using the miter structure and encoded as an oracle to be suitable for Grover's algorithm as a first straightforward approach. 
But this leads to pitfalls like the improper number of Grover iterations. 
The proposed approach provides a solution to this problem by iteratively evaluating different numbers of Grover iterations until a sufficiently distinct separation between possible counter examples and remaining states is determined.
Experimental evaluations confirmed the applicability of the approach: For all evaluation problem instances, the correct solution---either equivalence or non-equivalence with the corresponding counter examples---could be determined.
By this, this work provides a working concept that was able to classify all considered problem instances correctly.
Future work includes a more thorough empirical and/or theoretical evaluation---particular on the effect of the number of shots and the parameter $\phi$. 
\vspace{10cm}

\clearpage
\section*{Acknowledgments}
 This work received funding from the European Research Council (ERC) under the European Union’s Horizon 2020 research and innovation program (grant agreement No. 101001318), was part of the Munich Quantum Valley, which is supported by the Bavarian state government with funds from the Hightech Agenda Bayern Plus, and has been supported by the BMWK on the basis of a decision by the German Bundestag through project QuaST, as well as by the BMK, BMDW, the State of Upper Austria in the frame of the COMET program, and the QuantumReady project within Quantum Austria (managed by the FFG).

\printbibliography

\end{document}